\documentstyle[twocolumn,prb,aps,epsfig]{revtex}
\begin{document}
\draft
\title{Structural and electrical transport properties of superconducting
Au$_{0.7}$In$_{0.3}$ films: \\ A random array of
superconductor\--normal metal\--superconductor (SNS) Josephson junctions}
\twocolumn[
\hsize\textwidth\columnwidth\hsize\csname@twocolumnfalse\endcsname
\author{Yu. Zadorozhny and Y. Liu}
\address{Department of Physics, The Pennsylvania State University,
University Park, PA 16802}
\date{\today}
\maketitle

\begin{abstract}

The structural and superconducting properties of  Au$_{0.7}$In$_{0.3}$ films, grown by interdiffusion of alternating Au and In layers, have been studied.  The films were found to consist of a uniform solid solution of Au$_{0.9}$In$_{0.1}$, with excess In precipitated in the form of In-rich grains of various Au-In phases (with distinct atomic compositions), including intermetallic compounds. As the temperature was lowered, these individual grains became superconducting at a particular transition temperature ($T_c$), determined primarily by the atomic composition of the grain, before a fully superconducting state of zero resistance was established.  From the observed onset $T_c$, it was inferred that up to three different superconducting phases could have formed in these Au$_{0.7}$In$_{0.3}$ films, all of which were embedded in a uniform Au$_{0.9}$In$_{0.1}$ matrix.  Among these phases, the $T_c$ of a particular one, 0.8 K, is higher than any previously reported for the Au-In system.  The electrical transport properties were studied down to low temperatures.  The transport results were found to be well correlated with those of the structural studies.  The present work suggests that Au$_{0.7}$In$_{0.3}$ can be modeled as a random array of superconductor\--normal metal\--superconductor (SNS) Josephson junctions.  The effect of disorder and the nature of the superconducting transition in these Au$_{0.7}$In$_{0.3}$ films are discussed.

\end{abstract}

\pacs{74.40.+k,74.76.-w,74.76.Db}
]
\narrowtext

\section{Introduction} \label{intro}

The interplay between superconductivity and carrier localization has long
been a focus of research on disordered superconductors.  This question was
initially raised in the study of granular superconducting films.
\cite{granular} Such films, in which individual grains are separated by
insulating barriers, can be modeled as
superconductor-insulator-superconductor (SIS) tunnel junction
arrays.\cite{GoldmanBook} At lower temperatures, the film properties are
dictated by the competition between the inter-grain Josephson coupling and
the carrier confinement due to the charging energy of the junctions.  For
a superconducting grain, the number of Cooper pairs and the phase of the
superconducting order parameter are conjugate variables subject to an
uncertainty relation. \cite{Tinkham} As the fluctuation in the number of
Cooper pairs is suppressed due to the charging energy, the fluctuation in
the superconducting phase is enhanced. Josephson coupling between adjacent
superconducting grains, if strong enough, will lead to global phase
coherence.\cite{Goldman}

The localization effect has also been examined for homogeneously
disordered systems, such as ultrathin amorphous films, \cite{amorphous} in
which electrons in the normal state are strongly localized.  In this case,
superconductivity can also be suppressed, with a reduction in the
superconducting transition temperature or a complete loss of
superconductivity. \cite{Goldman} The same phase-number uncertainty
relation can be used to understand this phenomenon, since the fluctuation
in the number of electrons is again suppressed, due to localization.  In
fact, a theoretical model based on this picture has been suggested.
\cite{Zhou} As the amount of disorder increases, a structurally and
chemically homogeneous superconducting film will break down into
superconducting droplets, linked by Josephson coupling.  The
superconducting state will survive only if the coupling is sufficiently
strong.  This model makes it possible to understand the interplay between
superconductivity and localization in both granular and homogeneous films
on the same footing.

A natural question to ask is whether it is possible to identify a system
that can be disordered for superconductivity, but in which electrons or
Cooper pairs are not particularly localized.  An example of such a system
could be a granular superconductor with a
superconductor\--normal metal\--superconductor (SNS) type of intergrain
Josephson coupling. In this regard, it is interesting to note that, while
artificial proximity-coupled Josephson junction arrays\cite{jja} have been
examined in detail, including finite size effects \cite{Lobb} and the
effect of disorder, \cite{Garland} granular materials that can be modeled
as an SNS junction array, such as niobium nitride films, \cite{NbN} have
hardly been studied.

We have studied films of Au$_{0.7}$In$_{0.3}$.  These films are strongly
disordered as far as superconductivity is concerned due to a spatially
varying local superconducting transition temperature, $T_c^0$.  (Phase
separation in this binary alloy system is directly responsible for the
spatially varying $T_c^0$, as shown below.)  On the other hand, Cooper
pairs are not confined as in the case of conventional SIS granular films.
We report several generic features of these Au$_{0.7}$In$_{0.3}$ films,
grown at ambient temperature, in which the interdiffusion of Au and In led
to the formation of several superconducting Au-In minority phases,
including solid solutions and intermetallic compounds with distinct
concentrations of Au and In. We show that these Au-In films may be
regarded as random arrays of SNS junctions with unique superconducting
properties.  The morphology, chemical composition, and the low-temperature
electrical transport and superconducting properties of the system, which
were found to be well correlated, are documented.

The rest of the paper is organized as follows:  in Sec.~\ref{au-in} we
review various phases present in bulk Au-In alloy and summarize the
results on superconductivity in the Au-In binary system previously
reported in literature.  In Sec.~\ref{sample} we describe the techniques
used in film growth and characterization.  Sec.~\ref{structure} and
\ref{composition} deal with the film morphology and chemical composition.
Sec.~\ref{transport} and ~\ref{transport2} contain results on the
electrical transport and superconducting properties of these films.
Finally, Sec.~\ref{summary} summarizes the results.

\section{A\lowercase{u}-I\lowercase{n} Alloy} \label{au-in}

The phase diagram of Au-In binary alloy is rich and relatively well
studied. \cite{alloy} In the bulk form, the following alloy or
intermetallic compound phases, stable at room temperature, have been
reported \cite{alloy}: 1) face-centered cubic (fcc) solid solution with
room temperature solubility ranging from 0 to approximately 10 atomic \%
of In ($\alpha$ phase); 2) the Nd- or Ni$_3$Ti-type hexagonal $\alpha_1$
phase (11-13\% of In).  However, the existence of this phase as a stable
one was questioned \cite{alloy}; 3)  close-packed hexagonal (cph) solid
solution ($\zeta$, 10-16\% of In); 4-6) solid solution phases of
$\beta_1$, $\epsilon'$ and $\gamma'$, with In concentrations around 22\%,
25\%, and 30\%, respectively.  The structure and precise stoichiometry of
these phases are yet to be clarified\cite{alloy}; 7-8) intermetallic
compounds AuIn (triclinic) and AuIn$_2$ (of fluorite-type crystal
structure) with a negligible solubility of In or Au in these compounds.
There is no appreciable solid solubility of Au in In.  A schematic phase
diagram around the room temperature is shown in Fig.~\ref{1}, in which the
$\alpha_1$ phase is marked with a dashed line to reflect the uncertainty
associated with this phase. In thin-film samples prepared by
interdiffusion of Au and In at room temperature, a technique employed in
the present experiment, additional phases may be present.

Superconductivity in several phases of Au-In alloy has been reported, as
summarized in Ref.~\ref{alloy}.  In the bulk form, pure In has a
transition temperature of 3.40~K.  The $T_c$ of intermetallic compound
AuIn was reported to be between 0.4 and 0.6~K. \cite{auin} However, the
measurements were done on a powder ingot; the phase composition of the
sample was not independently verified.  Measurements on AuIn$_2$ ingots
yielded values of $T_c$ from 0.1~K \cite{auin2} to 0.2~K.\cite{NMR} For
the $\alpha$-phase, the $T_c$ of Au$_{0.9}$In$_{0.1}$, close to the limit
of solid solubility of In in Au, was 77~mK. \cite{Buchal} It further
decreased to 11~mK for Au$_{0.94}$In$_{0.06}$ \cite{Hoyt} and to 0.5~mK
for Au$_{0.98}$In$_{0.02}$. \cite{Buchal} The $T_c$ of the $\zeta$ phase
has been reported to be between 35 and 100~mK for a nominal composition
of Au$_{0.8}$In$_{0.2}$. \cite{Luo}

From these previous studies it is clear that there exist numerous
superconducting phases in the Au-In alloy system. In addition, a
significant range of In concentrations does not correspond to any single
equilibrium phase at room temperature.  As a result, depending on the
composition and the preparation technique, various phase-separated
samples are expected. Since a wide range of superconducting transition
temperature are possible for these phases, the local $T_c$ is expected to
vary spatially as well.

The Au-In system has been of great interest for both fundamental
research and practical applications.  In particular, it has served as a
model system for experimental studies of bulk and film solid diffusion and
compound formation. \cite{Finstad,Simic,Seyffert,Millares} The
$\alpha$-phase alloy attracted a large amount of attention in the pursuit
of possible superconductivity in pure Au. \cite{Buchal,Hoyt} Nuclear
ferromagnetism was recently found to coexist with superconductivity in
AuIn$_2$ at micro-kelvin temperatures. \cite{NMR2} We have previously
reported observations of strongly enhanced magnetic fingerprints,
\cite{MFP} and of fractional-flux Little-Parks oscillations in
disordered mesoscopic Au$_{0.7}$In$_{0.3}$ cylinders. \cite{splitting}

Au-In alloy is also important for practical applications.  The wide range
of superconducting transition temperatures in the same material system is
useful for applications where a well targeted $T_c$ is needed. Au-In films
are easy to prepare, adhere well to most substrates, and are structurally
stable. They do not oxidize significantly, either. AuIn$_2$ is used in
commercial superconducting quantum interference devices (SQUIDs).

\section{Sample Preparation and Measurements} \label{sample}

Au-In films were grown in a high vacuum chamber by sequential
deposition of 99.9999\% pure Au and In.  Glazed alumina or polished quartz
was used as substrates.  Three or five alternating layers were deposited,
the top and the bottom layers being Au.  The respective layer thicknesses
for Au and In, not exceeding 10~nm each, were determined based on the
desired stoichiometric ratio.  During evaporation, the deposition rate was
controlled using a quartz crystal thickness monitor.  A typical film was
6~mm long and 0.5~mm wide, patterned for 4-point electrical measurements.

The initial mixing of Au and In apparently occurred during the evaporation
at the ambient temperature.  Subsequently, interdiffusion continued at
room temperature.  Based on a previous study, the film structure should
have stabilized within several days. \cite{Simic} Depth profiles of
Au$_{0.7}$In$_{0.3}$ films revealed signs of incomplete interdiffusion of
Au and In right after the deposition. However, they disappeared after a
few days (Sec.~\ref{composition}).  Resistance measurements showed no
evidence for a superconducting transition at 3.4~K in any film, indicating
that no macroscopic fraction of pure In was present.

The surface of every Au-In film was studied by atomic force microscopy
(AFM) and/or by scanning electron microscopy (SEM).  X-ray photo-electron
spectroscopy (XPS) was used to probe the chemical composition of select
samples.  The relative concentrations of Au and In were inferred from the
energy spectra of emitted electrons, averaged over a surface area of the
order of 1~mm$^2$ and a depth of about 5~nm.  After each scan, the surface
layer was removed by Ar$^+$ ion etching, exposing the underlying material.
A depth profile of the chemical composition was generated by repeating this
procedure until all of the film was etched away.  Laterally resolved,
depth-averaged composition analysis was carried out using back-scattered
electron micrography (BSEM) and characteristic X-ray analysis.  Crystalline
structure of the films was also analyzed by X-ray diffractometry (XRD).

Electrical transport measurements of Au-In films were carried out in a
$^3$He or a dilution refrigerator equipped with a superconducting magnet.
All electrical leads entering the cryostat were filtered with the
attenuation of 10~dB at 10~MHz and 50~dB at 300~MHz.  The 4-point
resistance of the films was measured using a d.c. current source and a
digital nanovoltmeter.  The measuring current was
chosen in the ohmic regime of the current-voltage (I-V) characteristic
(typically $1 \mu $A).

\section{Film Morphology} \label{structure}

It is known that film morphology for a particular material system can vary
widely depending on the preparation condition. The film morphology will,
in general, affect the electrical transport and superconducting
properties of the film. Therefore it is important to document the general
morphological features of Au$_{0.7}$In$_{0.3}$ films.  It was found that,
depending on the evaporation rate ({\it e}) and the total thickness ({\it
t}), two different types of Au$_{0.7}$In$_{0.3}$ films, marked by its
distinct surface morphology, could be grown.

Films thinner than 30~nm were found to be smooth regardless of the
evaporation rate ($0.1 \leq e \leq 2$~nm/s).  Thick films ($t \geq
30$~nm) were also smooth if deposited at about 0.5~nm/s or slower.
However, micron-size clusters were found in thick films ($t \geq 30$~nm)
that had been deposited at 1-2~nm/s (Sec.~\ref{structure}), resulting in
bumpy films. The change in the film morphology for thicker films may be
attributed to the higher atomic mobility during fast deposition.  In
thinner films, the formation of clusters was suppressed probably because
of the reduced atom mobility originating from substrate effects.

AFM pictures of two Au$_{0.7}$In$_{0.3}$ films are shown in Fig.~\ref{2}.
Film~\#25, nominally 10~nm thick, represents the smooth films.
The film was indeed uniform at large length scales, as seen in
Fig.~\ref{2}a.  A close-up view (Fig.~\ref{2}b) revealed the presence of
grains with a typical size of 20~nm, larger than the nominal film
thickness.  This may indicate that the film was formed by a single layer
of grains.  The grains were most likely crystalline, according to XRD
studies of thicker Au$_{0.7}$In$_{0.3}$ films.

A 40~nm thick sample, film~\#29 shown in Figs.~\ref{2}c and d, is
representative of bumpy films, was prepared by depositing at a rate faster
than 1~nm/s. For these films, the variation in the surface height was
comparable to the total film thickness.  Distinct clusters, showing up as
bright spots in the picture, protruded from the rest of the film surface
by about 20~nm.  The clusters, several microns apart, often appeared to
have a somewhat rounded hexagonal shape, with a typical size of 1~$\mu $m.
The clusters were surrounded by circular terraces, which were 4-5~$\mu$m
in diameter and elevated above the rest of the film by a few nm.

A closer examination (Fig.~\ref{2}d) revealed that the size of the grains
in the clusters is approximately 60~nm, while the grains comprising
surrounding areas were similar in size to those in uniform films. This
probably was either a consequence of the larger film thickness at the
protruding clusters, or an indication that these clusters were of a Au-In
phase different from that of the surrounding area.

\section{Phase Composition} \label{composition}

Another important set of questions is related to the spatial variation of
the chemical composition of the Au$_{0.7}$In$_{0.3}$ films, and whether it
correlates with the film morphology.  To answer these questions, depth
profiling of the chemical composition was carried out by XPS.  The XPS
analysis yielded direct information on the chemical composition in the
vertical direction of the film.  When combined with AFM imaging, the XPS
depth profiling provided information on chemical composition in the
lateral direction, as shown below.

A 100~nm thick sample (film~\#28) was prepared by alternating evaporation
of six 10~nm thick layers of Au and five 8~nm thick layers of In onto a
quartz substrate.  An XPS depth profiling analysis was carried out
immediately after the film deposition. Fig.~\ref{3}a contains an
energy spectrum of photo-electrons emitted from the film's original
surface.  Au and In concentrations were calculated from the areas of
Au~4p$_{3/2}$ and In~3d$_{5/2}$ lines, using standard sensitivity factors
of 3.8 and 0.62 respectively.  Traces of oxygen and carbon were found at
the surface, likely due to post-deposition contamination in the air,
but disappeared deeper into the film.

Shown in Fig.~\ref{3}b are two depth profiles of film~\#28, obtained by
taking XPS scans at approximately 4~nm steps.  For the profile taken
immediately after the deposition, remnants of the layers are evident near
the surface but become less distinct as the substrate is approached.
This can be understood by noting that the bottom layers, deposited first,
were exposed to elevated temperatures during evaporation longer and
therefore had interdiffused more. In the depth profile obtained
after the film had been aged, no evidence of any layered structure was
found.  The measured In concentration was approximately 28\% at the
surface of the film and 10\% deep in the film.

The above observation can be understood by taking into account the rate of
ion etching for different alloy phases.  In Fig.~\ref{4}, we show a
sequence of AFM scans of the film's surface at different stages of
profiling.  The surface roughness increased significantly after etching,
suggesting that during ion etching, the In-rich grains were preferentially
sputtered away.  Indeed, after two or three etching periods the XPS could
only detect the Au$_{0.9}$In$_{0.1}$ component, in agreement with this
preferential etching picture. Taking into account the bulk phase diagram
described above, it is clear that the uniform matrix was the saturated
solid solution, Au$_{0.9}$In$_{0.1}$ $\alpha $ phase. An XPS scan of the
original film surface yielded In concentration close to the true average
value of 28\%.  The difference between the measured In concentration and
the expected Au$_{0.7}$In$_{0.3}$ composition at the film surface is most
likely due to the incomplete interdiffusion of Au and In in the surface
layer immediately after the deposition, with Au being the ending layer.

The electrical transport measurements, discussed below, suggested that the
In-rich grains found in the Au$_{0.7}$In$_{0.3}$ films might be made of
different phases, including intermetallic compound AuIn.  A previous study
of the room temperature interdiffusion of Au and In in Au-In bilayers
\cite{Simic} showed that for In fraction between 23 and 36\%, both AuIn
and AuIn$_2$ compounds were formed initially in the excess Au.  However,
after several days, and for months thereafter only AuIn could be found.
If the same pattern of phase formation was applicable to our films as
well, then the In-rich grains might be predominantly intermetallic
compound AuIn.

The presence of protruding clusters in the bumpy films is very striking.
Could the chemical composition of these clusters
be different from the rest of the film?  XPS studies of these bumpy
films yielded depth profiles similar to those described above.  Most
importantly, the protruding clusters were not preferentially etched.
Therefore these clusters were likely {\it not} the same In-rich phases
found in the smooth films.  The AFM images away from the clusters showed
similar features (Fig.~\ref{4}), suggesting that away from the
clusters the film had similar compositional
distribution as in the smooth films. While back-scattered electron
micrography (BSEM), characteristic X-ray analysis, and X-ray
diffractometry (XRD) were all attempted, the precise chemical composition
of the protruding clusters remain undetermined, mainly due to the lack of
reliable standards.  To determine the precise phase of these clusters,
considerable more work on crystallography of the Au-In system, including
the preparation of pure phase samples to be used as standards, would be
needed.  These experiments, which are beyond the scope of the present
study, were not carried out.

The picture emerging from these morphological and chemical studies is one
in which multiple material phases were formed in the films.  The
predominate phase appears to be the saturated Au$_{0.9}$In$_{0.1}$
$\alpha$ phase.  The others were various In-rich phases.  The
Au$_{0.9}$In$_{0.1}$ formed a matrix with In-rich regions embedded in it
randomly, in both the position and the size.

\section{Electrical transport properties of
A\lowercase{u}$_{0.7}$I\lowercase{n}$_{0.3}$}
\label{transport}

The characteristic features in film morphology and chemical composition
are expected to dictate the electrical transport properties of the films.
In Fig.~\ref{5} we show normalized resistances plotted against temperature
for two sets of Au$_{0.7}$In$_{0.3}$ films of varying thickness.  In each
set, the films were grown under the same nominal conditions.  The films in
Fig.~\ref{5}a were prepared by slow deposition, with all films
exhibiting a smooth morphology, free of the clusters.  The films shown in
Fig.~\ref{5}b were prepared by fast deposition. Protruding
clusters such as those shown in Fig. 2c and d appeared clearly for films
with thickness of 30~nm and larger.

For films of multiple phases, the onset of the superconducting transition
usually marks the highest local $T_c$ for a particular phase.  Films
prepared by slow deposition shown in Fig.~\ref{5}a had three onset
superconducting transition temperatures, roughly 0.45K, 0.65K, and 0.8K,
depending on film thickness.  On the other hand, for films prepared by
fast deposition shown in Fig. \ref{5}b, the onset of superconductivity was
found at 0.45 and 0.8K only, with a sharp drop in film resistance at
0.6~K. These onset temperatures probably correspond
to the $T_c$ of different In-rich phases.

The temperature, 0.65K, is consistent with the highest $T_c$ reported for
the intermetallic compound AuIn.  Combining results obtained in
morphological and compositional analyses presented above, it is
reasonable to assume that this In-rich phase was AuIn.  However, exactly
which two phases were responsible for superconductivity at 0.45 and 0.8K
are more difficult to determine.  In this regard, it should be noted that
0.8K is higher than any superconducting transition
temperature previously reported for the Au-In system.  The $T_c$ reported
for saturated Au$_{0.9}$In$_{0.1}$ $\alpha $ phase was 77~mK.

Two smooth films, 30 and 40nm thick and prepared by fast deposition,
showed two sharp drops in film resistance around 0.3 and 0.6K respectively
(Fig.~\ref{5}a).  The drop at 0.6K should correspond to the
superconducting transition of grains of compound AuIn. In principle, the
second drop in resistance could correspond to the superconducting
transition of a percolation path formed by the phase with a local $T_c$
around 0.45K as identified above, or it might be due to emergence of a
global phase coherence among the AuIn grains, which were responsible for
the resistance drop at 0.6K.

Experimentally, it was found that the temperature of this second drop
could be increased significantly after low-temperature ($< 400$~K)
annealing by exposing the sample to a heat lamp for one hour in vacuum
(Fig.~\ref{6}). Based on the bulk phase diagram of Au-In alloy, such
low-temperature annealing should not lead to the formation of a new stable
phase.  Indeed, the temperature corresponding to the first resistance drop
remained essentially unchanged after annealing.  On the other hand, the
normal state resistance was reduced significantly. It is known the
Josephson coupling in an SNS junction depends on the {\it N}-layer
resistance.  Therefore, it appears that the formation of a separate
superconducting phase was not the cause for the resistance drop found at
the lower temperature.

Two-step transitions were observed previously in proximity-coupled
Josephson junction arrays. \cite{jja} However, because the local $T_c$ of
the superconducting islands is uniform and equal to the bulk $T_c$ of the
superconductor, the resistance dropped sharply at this temperature. In the
present study, because of the proximity effect, the In-rich grains
embedded in the Au$_{0.9}$In$_{0.1}$ $\alpha $ phase matrix (which is a
normal metal above its $T_c$ of 77mK) would have a local $T_c$ dependent
on their size, and the exact local In concentration.  Below the
superconducting transition of these In-rich grains, but above that of
Au$_{0.9}$In$_{0.1}$, the films should behave as random networks of SNS
(proximity-coupled) Josephson junctions. As temperature is lowered, a
sharper transition to a zero-resistance superconducting state was observed
due to the Josephson coupling of all superconducting grains. This picture
is fully consistent with the one inferred from the structural and
compositional analyses.

\section{Effect of Disorder and the Nature of Superconducting Transition}
\label{transport2}

The normal-state resistivity of a Au$_{0.7}$In$_{0.3}$ film is a good
measure of the degree of disorder that affects electron motion in the
normal state.  The normal-state resistivity can be controlled by the film
thickness.  In the present study, the underlying phases of films prepared
by slow deposition may have been affected strongly by the details of the
substrate conditions. On the other hand, the films prepared by fast
deposition shown in Fig.~\ref{5}b revealed behavior as a function of film
thickness more systematic than those prepared by slow deposition, and
therefore will be the focus of this discussion.

The onset of the superconducting transitions of these films prepared by
fast deposition was found to be at either 0.45K or 0.8K, corresponding to
two possible Au-In phases,  rather than three, as seen in the case of
films prepared by the slow deposition. The zero-resistance $T_c$ corresponding
to the onset of the global phase coherence was suppressed from about 0.6~K
for the 35-40~nm thick films to about 0.2~K for the 8~nm thick film.  For
the 5~nm thick film, no superconductivity was found.  However, even such a
thin film remained metallic down to 0.3~K, indicating that the electrons
in Au$_{0.7}$In$_{0.3}$ films are delocalized in the normal state.

The normal state sheet resistance, while changing from a few to a few
dozen ohms (see Table~\ref{tab1}), remained very low for all films
studied.  For thicker films, the values of the sheet resistance scaled
with the film thickness, yielding a resistivity of approximately $12 \cdot
10^{-6}~\Omega \cdot$cm at 1~K.  Despite of the small normal-state
resistivity, thin Au$_{0.7}$In$_{0.3}$ films had a broad transition regime
($> 0.2$~K), as seen in Fig. \ref{5}b. Such a broad transition is
typically found in strongly disordered superconductors due to strong phase
fluctuations.\cite{amorphous} In artificially prepared SNS junction
arrays, the transition width was found to increase with the increase of
disorder.\cite{Garland} The broad superconducting transition is a clear
indication that Cooper pairs experience substantial disorder despite the
low normal-state resistance.

The relatively thick Au$_{0.7}$In$_{0.3}$ films, such as the 30nm and 35nm
thick films, showed a distinctly different behavior.  First, there was a
gradual decrease in film resistance right below the onset $T_c = 0.8$~K.
As the temperature was lowered, sharp drop to zero resistance was found at
$T=0.6$~K (slightly below the local $T_c$ of the compound AuIn, 0.65K).
The temperature dependence of the resistance may be qualitatively
explained using the following picture: The local $T_c$ of the protruding
clusters is 0.8~K. These clusters induced superconductivity in their
surrounding regions below 0.8~K due to proximity effect.  The size of
these superconducting regions increased as the decreasing temperature,
resulting in a gradual drop in film resistance.  Finally, slightly below
0.6K, a zero-resistance superconducting state was established due to
either the emergence of global phase coherence in an SNS junction array
formed by the protruding clusters, or alternatively by the formation of a
percolation path of grains of another superconducting phase (see below for
further discussion), such as the compound AuIn with a onset $T_c$ of 0.6K.
In either case, even the relatively thick Au$_{0.7}$In$_{0.3}$ films with
a sharp transition to global superconductivity may possess a high degree
of disorder for Cooper pairs because of the inhomogeneity in the amplitude
of superconducting order parameter, even though the normal state is a
perfectly good metal, making these Au$_{0.7}$In$_{0.3}$ films a unique
superconducting system.

In Fig.~\ref{8}, the results of more detailed resistance measurements are
presented for 30~nm thick film.  Perpendicular ($B_\perp$) and parallel
($B_{||}$) magnetic fields were found to affect the superconducting
transition of the film in very different ways.  This was expected since
$B_\perp$ created superconducting vortices in the film, which would induce
a finite resistance if they were depinned and started to creep in the
film. On the other hand, the effect of $B_{||}$ was primarily to suppress
superconductivity of each individual grain via the Zeeman effect, as in
this case free vortices cannot form in the film. Correspondingly, a much
lower $B_\perp$ than $B_{||}$ was needed to suppress the global
superconducting transition.

The motion of vortices in an SNS Josephson junction array is related
directly to phase slips.  Enhanced phase slips, and therefore phase
fluctuations, will lead to a wide superconducting transition regime and a
suppressed global $T_c$, as shown in Fig.~\ref{9}.  Similar behavior was
found when the film was made thinner, as discussed above.  In a parallel
field, while the field does not introduce free vortices to the film
directly, suppressing superconductivity in the grains with a sufficiently
large $B_{||}$ will lead to enhanced phase slip, which involves the
unbinding of a vortex-antivortex pair, again resulting in a wider
superconducting transition regime.  This effect is illustrated clearly in
Fig.~\ref{9}, where the magnetic field dependence for both on the global
and the onset (local) $T_c$ is shown.

It is interesting to note that the resistance just above the drop, $R_p$,
corresponding to either the normal-state resistance of the SNS junction
array or the resistance of the link or links that complete the
percolating path, showed an intriguing magnetic field dependence. It was
found that $R_p$ varied smoothly with the amplitude of the field,
independent of the field orientation, as shown in Fig.~\ref{10}.  In
addition, in small perpendicular magnetic fields, $R_p$ increased with
field (Fig.~\ref{10}).  Such a variation may provide a hint as to how the
global phase coherence emerges in these bumpy films. To clarify such field
dependence of $R_p$, more theoretical and experimental work is needed.

\section{Conclusion} \label{summary}

In conclusion, we have presented results on the study of
superconducting films of Au$_{0.7}$In$_{0.3}$ alloy with varying
thickness. The films were grown by sequential deposition and
interdiffusion of Au and In.  The structure and the electrical transport
studies of Au$_{0.7}$In$_{0.3}$ films, especially the observations of the
onset superconducting transition temperatures of the films, suggest the
presence of several Au-In phases, including, most likely,
Au$_{0.9}$In$_{0.1}$ ($T_c = 77$~mK), two In-rich phases ($T_c =0.45$ and
0.65~K, respectively), and the intermetallic compound AuIn ($T_c=0.6$~K).

At temperatures below their onset superconducting transition temperature,
the In-rich regions underwent a superconducting transition and became
Josephson coupled with one another, resulting in a random network of
Josephson junctions of SNS type, similar to previous measurements of
artificial arrays of SNS Josephson junctions. The present work appears to
have established a unique system which is disordered as far as the
Cooper paris are concerned, while the normal state remains a good
metal.

\section*{Acknowledgments}

The authors would also like to thank Vince Bojan, Mark Angelone, Tom
Rusnak, and Ed Basgall of Penn State Materials Characterization Laboratory
for help with characterization of the Au$_{0.7}$In$_{0.3}$ films.  They
would like to thank Mari-Anne Rosario for discussion and help with
preparing the manuscript.  This work is supported by NSF under grant
DMR-9702661.

\begin{table}
\caption{Summary of the sample characteristics of Au$_{0.7}$In$_{0.3}$
films discussed in this paper, including morphology ("s" for smooth films,
"b" for bumpy films with protruding clusters), nominal film thickness $d$,
evaporation rate $e$, normal-state sheet
resistance $R_N^\Box$, onset temperature for superconductivity $T_c^{\rm
onset}$, and zero-resistance transition temperature $T_c$.  Actual
thickness, estimated from AFM edge profiles for some films, was larger than
$d$ by several percent.  $T_c^{\rm onset}$ and $T_c$ were evaluated at the
$0.99R_N$ and $0.01R_N$ resistance level, respectively. Electrical
transport measurements were not carried out for Films \#28,
29, and 32, which were prepared for structural and compositional studies.}
\label{tab1}
\medskip
\begin{tabular}{ddddddd}
film & s/b & $d$,~nm & $e$,~nm/s  &$R_N^\Box,~\Omega$ & $T_c^{\rm onset}$,
K & $T_c$, K \\
\tableline
\#8	& b	& 30	& 1.2	& 4.2	& 0.734	& 0.586	\\
\#9	& s	& 20	& 0.9	& 11.5	& 0.669	& 0.378	\\
\#10	& b	& 35	& 1.6	& 3.9	& 0.804	& 0.561	\\
\#11	& s	& 25	& 1.0	& 6.4	& 0.767	& 0.561	\\
\#13	& s	& 28	& 1.8	& 6.6	& 0.631	& 0.375	\\
\#16	& s	& 15	& 1.0	& 34.4	& 0.423	& 0.333	\\
\#18	& s	&  8	& 0.5	& 89.1	& 0.431	& 0.229	\\
\#19	& s	&  5	& 0.5	& 3000	& \--	& \--	\\
\#21	& s	& 40	& 0.4	& 3.1	& 0.624	& 0.321	\\
\#22	& s	& 30	& 0.4	& 4.7	& 0.653	& 0.340	\\
\#23	& s	& 20	& 0.5	& 8.2	& 0.791	& 0.374	\\
\#24	& s	& 10	& 0.3	& 25.0	& 0.456	& 0.300	\\
\#25	& s	&  8	& 0.3	& 42.6	& 0.416	& 0.273	\\
\#28	& s	&100	& 0.6	& \--	& \--	& \--	\\
\#29	& b	& 40	& 1.3	& \--	& \--	& \--	\\
\#32	& b	&100	& 1.5	& \--	& \--	& \--	\\
\end{tabular}
\end{table}

\begin{figure}
%\centerline{\epsfig{file=1.eps,angle=0,width=3.4in}}
\medskip
\caption{Schematic Au-In phase diagram, after Ref.~\ref{alloy}.  Phases
reported to be superconducting include pure In ($T_c=3.4$~K), AuIn$_2$
($T_c=0.1-0.2$~K), AuIn ($T_c=0.4-0.6$~K), $\zeta$ ($T_c \approx
0.1$~K), and $\alpha$ ($T_c=0-77$~mK)}.
\label{1}
\end{figure}

\begin{figure}
%\centerline{\epsfig{file=2.eps,angle=0,width=3.4in}}
\medskip
\caption{(a-d)  AFM images of two Au$_{0.7}$In$_{0.3}$ films.  Brighter
areas correspond to more elevated regions. (a,b)  are for 10~nm thick
sample, film~\#25. Image areas are $20 \times 20$ and $1 \times
1~\mu$m$^2$ respectively. Image depth, the surface height variation
corresponding to the entire gray scale, is 5~nm for both pictures. (c,d)
are for 40~nm thick (film~\#29).  Image areas $20 \times 20$ and $2 \times
2~\mu$m$^2$, with an image depth of 100~nm.}
\label{2}
\end{figure}

\begin{figure}
%\centerline{\epsfig{file=3.eps,angle=0,width=3.4in}}
\medskip
\caption{(a)  Photoelectron spectrum of the surface layer of a 100~nm thick
Au$_{0.7}$In$_{0.3}$ sample, film~\#28.  Major lines are identified.
(b) Depth profiles of the relative concentrations of Au and In for
film~\#28.
The concentrations are plotted as functions of the etching time (see text),
which is proportional to the depth.  Open circles -- immediately after
deposition; solid circles -- 2 months after the initial analysis.}
\label{3}
\end{figure}

\begin{figure}
%\centerline{\epsfig{file=4.eps,angle=0,width=3.4in}}
\medskip
\caption{AFM images of film~\#28 at different stages of depth profiling:
(a) before ion etching, (b) with $\approx5$~nm removed, and (c) with
$\approx50$~nm removed.  All pictures are $10\times10~\mu$m$^2$, with
an image depth of 100~nm.  Preferential sputtering of the In-rich phase is
evidenced by the surface roughening in the process.}
\label{4}
\end{figure}

\begin{figure}
%\centerline{\epsfig{file=5.eps,angle=0,width=3.4in}}
\medskip
\caption{Normalized resistance plotted as a function of temperature for two
sets of Au$_{0.7}$In$_{0.3}$ films of varying thickness.
(a) Films grown by slow deposition.  The thicknesses are 8~nm (film~\#25),
10~nm (film~\#24), 20~nm (film~\#23), 30~nm (film~\#22), and 40~nm
(film~\#21).  The film parameters are shown in Table~1.
(b) Films grown by fast deposition.  The film thicknesses are 5~nm
(film~\#19), 8~nm (film~\#18), 15~nm (film~\#16), 20~nm (film~\#9), 25~nm
(film~\#11), 30~nm (film~\#8), and 35~nm (film~\#10), as indicated.  The
two thickest films contained protruding clusters.}
\label{5}
\end{figure}

\begin{figure}
%\centerline{\epsfig{file=5.eps,angle=0,width=3.4in}}
\medskip
\caption{Resistance plotted as a function of temperature for film~\#22
before (left curve) and after annealing (right curve).}
\label{6}
\end{figure}

\begin{figure}
%\centerline{\epsfig{file=5.eps,angle=0,width=3.4in}}
\medskip
\caption{Normal-state resistance, taken at $T=1$~K, plotted as a function
of inverse of the film thickness for films shown in Fig. 5.  The filled
circles are for films grown by slow deposition, while the empty circles
are for films prepared by fast deposition.}
\label{7}
\end{figure}

\begin{figure}
%\centerline{\epsfig{file=6.eps,angle=0,width=3.4in}}
\medskip
\caption{Resistance of 30~nm thick bumpy film (film \#8), plotted as
a function of temperature for different values of (a) perpendicular and
(b) parallel magnetic fields, and as a function of (c) perpendicular and
(d) parallel field at different temperatures.  For (a), from
top to bottom, $B_{\bot } = 300,$ 170, 160, 150, 120, 110, 100,
75, 50, 25, 10, and 0~G. For (b), from top to bottom, $B_{\| } =
4900,$ 3920, 2940, 2450, 2960, 1500, 980, and 0~G.  For (c), from
top to bottom, $T = 0.65,$ 0.625, 0.60, 0.55, 0.50, 0.40, and
0.33~K.  For (d), from top to bottom, $T = 0.60,$ 0.55, 0.50, 0.425,
and 0.33~K. Measurements were carried out during two separate cool-downs.
Temperatures below 0.3~K were not available for these measurements.}
\label{8}
\end{figure}

\begin{figure}
%\centerline{\epsfig{file=5.eps,angle=0,width=3.4in}}
\medskip
\caption{Values of the onset and the global $T_c$ plotted
as a function of magnetic field in (a) perpendicular and (b) parallel
directions, for film~\#8.}
\label{9}
\end{figure}

\begin{figure}
%\centerline{\epsfig{file=5.eps,angle=0,width=3.4in}}
\medskip
\caption{Critical resistance, $R_p$, plotted as a function of magnetic
field in perpendicular (open circles) and parallel (filled circles)
directions for film~\#8.}
\label{10}
\end{figure}

\end{document}